\documentclass[12pt]{article}
\usepackage[noblocks]{authblk}
\usepackage{here}
\usepackage{subfig}
\usepackage{graphicx}
\usepackage{color}
\usepackage{braket}
\usepackage{amssymb, amsmath}
\usepackage{url}

\usepackage{hyperref}
\usepackage[samesize]{cancel}
\usepackage{fancyhdr}

\newcommand{\vect}[1]{\mbox{\boldmath${#1}$}}
 
\newcommand{\be}{\begin{eqnarray}}
\newcommand{\ee}{\end{eqnarray}}

\newcommand{\beq}{\begin{equation}}
\newcommand{\eeq}{\end{equation}}
\newcommand{\beqa}{\begin{eqnarray}}
\newcommand{\eeqa}{\end{eqnarray}}

\newcommand{\lmk}{\left(}
\newcommand{\rmk}{\right)}
\newcommand{\lkk}{\left[}
\newcommand{\rkk}{\right]}
\newcommand{\lnk}{\left\{}
\newcommand{\rnk}{\right\}}

\newcommand{\la}{\langle}
\newcommand{\ra}{\rangle}

\newcommand{\vs}{{\vect s}}
\newcommand{\vx}{{\vect x}}
\newcommand{\vy}{{\vect y}}
\newcommand{\vw}{{\vect w}}

\def\nat{Nature}

\def\apj{Astrophysical Journal}
\def\apjs{Astrophysical Journal, Supplement}
\def\aj{Astronomical Journal}

\def\prd{Physical Review D}

\date{}
\title{\vspace{-0.5cm}\textbf{Application of the independent component analysis to the iKAGRA data} 
}
\author[1,2]{T.~Akutsu}
\author[3,4,1]{M.~Ando}
\author[5]{K.~Arai}
\author[5]{Y.~Arai}
\author[6]{S.~Araki}
\author[7]{A.~Araya}
\author[3]{N.~Aritomi}
\author[8]{H.~Asada}
\author[9,10]{Y.~Aso}
\author[11]{S.~Atsuta}
% ...
\author[　]{(KAGRA Collaboration)}

\affil[1]{Gravitational Wave Project Office, National Astronomical Observatory of Japan (NAOJ), Mitaka City, Tokyo 181-8588, Japan}
\affil[2]{Advanced Technology Center, National Astronomical Observatory of Japan (NAOJ), Mitaka City, Tokyo 181-8588, Japan}
\affil[3]{Department of Physics, The University of Tokyo, Bunkyo-ku, Tokyo 113-0033, Japan}
\affil[4]{Research Center for the Early Universe (RESCEU), The University of Tokyo, Bunkyo-ku, Tokyo 113-0033, Japan}
\affil[5]{Institute for Cosmic Ray Research (ICRR), KAGRA Observatory, The University of Tokyo, Kashiwa City, Chiba 277-8582, Japan}
\affil[6]{Accelerator Laboratory, High Energy Accelerator Research Organization (KEK), Tsukuba City, Ibaraki 305-0801, Japan}
\affil[7]{Earthquake Research Institute, The University of Tokyo, Bunkyo-ku, Tokyo 113-0032, Japan}
\affil[8]{Department of Mathematics and Physics, Hirosaki University, Hirosaki City, Aomori 036-8561, Japan}
\affil[9]{Kamioka Branch, National Astronomical Observatory of Japan (NAOJ), Kamioka-cho, Hida City, Gifu 506-1205, Japan}
\affil[10]{The Graduate University for Advanced Studies (SOKENDAI), Mitaka City, Tokyo 181-8588, Japan}
\affil[11]{Graduate School of Science and Technology, Tokyo Institute of Technology, Meguro-ku, Tokyo 152-8551, Japan}
% ...

\begin{document}

\maketitle

\begin{abstract}
We apply the independent component analysis (ICA) to the real data from a gravitational wave detector for the first time. 
Specifically we use the iKAGRA data taken in April 2016, and calculate the correlations between the gravitational wave strain channel and 35 physical environmental channels. Using a couple of seismic channels which are found to be strongly correlated with the strain, we perform ICA. Injecting a sinusoidal continuous signal in the strain channel, we find that ICA recovers correct parameters with enhanced signal-to-noise ratio, which demonstrates usefulness of this method.
Among the two implementations of ICA used here, we find the correlation method yields the optimal result for the case environmental noises act on the strain channel linearly.
\end{abstract}

\newpage

\section{Introduction}
Ever since Einstein found the existence of a gravitational wave solution in his theory of general relativity in
1916, it took exactly a century for mankind to succeed in its direct detection.  This delay is primarily 
due to the fact that the gravitational force is an exceedingly weak force compared with other interactions.

 The first detection of a gravitational wave by the advanced Laser Interferometer Gravitational wave Observatory (aLIGO) \cite{LIGO} brought a great impact on science and told the beginning of gravitational wave astronomy.
Following aLIGO and advanced Virgo, the large-scale cryogenic gravitational wave telescope (LCGT) now known as KAGRA, has been constructed in Kamioka, Japan \cite{KAGRA3}.
KAGRA will play very important roles in the international network of gravitational wave detection by measuring the number of polarization property, which is indispensable to prove the general relativity~\cite{Hagihara:2018azu}, and by improving the sky localization of each event significantly~\cite{Aasi:2013wya}. As the first underground and cryogenic detector, it will also provide important information to the third-generation detectors.
 
Because gravity is the weakest force among the four elementary interactions, gravitational waves have high penetrating power.
Therefore, unlike electromagnetic waves, they can propagate without being influenced by interstellar medium. In the same way, they enable us to see deep inside dense matter, such as the core of neutron stars, and bring information that electromagnetic waves cannot.
On the other hand, due to this property, gravitational wave signals tend to be quite small and its detection becomes very difficult. Thus, it is important to develop methods for extraction of these tiny signals.
There are a number of methods which extract signal out of large noises such as matched filtering~\cite{GAUSSIANmethod},
which yields an optimal result if (and only if) underlying noise is Gaussian distributed.
However, the problem is not so simple, as it is known that there exist non-Gaussian noises in real data.
They decrease the performance of the analysis methods assuming Gaussianity of the noises.
What is worse, these noises may be mistaken for true signals and increase the false alarm probability. 
Thus, it is necessary to deal with non-Gaussianity properly as stressed in \cite{first}.
Characterization, mitigation, and even subtraction of these noises in gravitational wave detector outputs 
have been extensively studied in the literature. The standard way including pre-data conditioning (whitening, band-passing), 
line-removal, and $\chi^2$ veto are well overviewed in~\cite{2019arXiv190811170T}. Many of recent works demonstrate performance of  
Deep Neural Networks \cite{2017CQGra..34f4003Z,2018CQGra..35i5016R,2018PhRvD..97j1501G,2018CQGra..35o5017P,2018PhRvD..98h4016P,2019PhRvD..99h2002C}, 
but see also \cite{2019arXiv190805644Z}.

In this situation, independent component analysis (ICA)~\cite{ICA1,ICA2,ICA3}
, which seperates mixed signal components, 
occupies a unique position among methods of signal processing because it makes use of non-Gaussianity of signals and noises instead of treating it as an obstacle. 
ICA has been used in various fields in astronomy, {\it e.g.}, \cite{MF_ICA, 2008A&A...491..597L, Ichiki_2013,2015ApJS..219....5Q, 2015MNRAS.447..400A, 2015ApJ...802..117M, 2016AJ....152...44I, 2016ApJ...820...86M, 2017Natur.548..555H}. For example, reference~\cite{MF_ICA} demonstrated ICA (EFICA and WASOBI) performance on simulated data mimicking two gravitational wave interferometer outputs.
The current paper, on the other hand, demonstrates it using real gravitational wave strain data from the iKAGRA detector and multiple real auxiliary channels that recorded status of the detector. 
ICA can separate various components obeying non-Gaussian distributions, so that it can remove (part of) non-Gaussian noises from strain data that records gravitational wave signals. Then the strain channel would consist of the real signal and (nearly) Gaussian noises.
Therefore ICA can support conventional matched filter technique as a non-Gaussian noise subtraction scheme.
In addition, ICA can be used even in the cases noises are nonlinearly coupled to the strain channel as demonstrated in~\cite{Morisaki:2016sxs}.

In this paper, we use the correlation method~\cite{Morisaki:2016sxs} (or 
the Gram-Schmidt orthogonalization method in the case of multiple channels in general) and FastICA~\cite{FastICA}.   
The former method, although conceptually different in derivation, has practically the same expression as the Wiener filtering~\cite{1999gr.qc.....9083A,2012RScI...83b4501D,2012CQGra..29u5008D,2014CQGra..31j5014M,2015CQGra..32p5014T,2019PhRvD..99d2001D} which has been used to analyze Caltech 40m and LIGO data, and quite remarkable success was recently reported~\cite{2019PhRvD..99d2001D} by using witness sensors including voltage monitors of analogue electronics for power main 
and photodiodes that monitor beam motion and its size for beam jitter.
We report the results of application of these two different ICA methods to the iKAGRA data and discuss its
usefulness in gravitational wave data analysis. 
The paper is organized as follows. In \S~\ref{ICA}, we introduce ICA in the simplest case where only one environmental channel is incorporated to the strain channel and review analytic formulas of correlation method obtained in our previous paper~\cite{Morisaki:2016sxs}. Then we extend this method to the case where two different environmental channels are concerned. We also introduce FastICA which is formulated in a different way.
In \S~\ref{result}, we present our application of ICA to the iKAGRA data with injected artificial continuous signal. Then we discuss the result focusing on the difference of the two methods in \S~\ref{discuss}. We argue that for the current setup where noises measured by the environmental channels affect the strain linearly and additively, the what we call correlation method yields the optimal result. The final section \S~\ref{conclusion} is devoted to conclusion.

\section{Independent Component Analysis (ICA)}\label{ICA}
As is seen in our previous paper~\cite{first}, signal detection under non-Gaussian noises is much more involved than the case with Gaussian noises since the optimal statistic has much complicated forms. ICA is an attractive method
of signal processing because it makes use of non-Gaussian nature of the
signals~\cite{ICA1,ICA2,ICA3} (see~\cite{icab1,icab2} for textbooks). We here introduce two methods of ICA as a method of non-Gasussian noise subtraction.

Basically, this method assumes only statistical independence between the signal and noises, and does not impose any other conditions on their distributions. 
However, a simpler formulation can be achieved by using physical information of gravitational wave detection as expressed in~\cite{Morisaki:2016sxs}. Following~\cite{Morisaki:2016sxs}, we first formulate the subtraction of non-Gaussian noise in the gravitational wave detection for the case where noise is linearly coupled to the strain. Then we introduce analytic formulas of ICA for this case, which we call the correlation method. Previously, this was two component analysis in~\cite{Morisaki:2016sxs}, but we here developed a multiple component version for combining different environmental channels.

On the other hand, there is a robust formulation which does not incorporate any information of the concerned system, which is called FastICA~\cite{FastICA}. We also introduce this method in this section and apply it in our analysis as a comparison. 

\subsection{Removing non-Gaussian noises}\label{problem}
In this paper, we consider the following simple problem as a first step to test applicability of ICA for detection of
GWs. Let us consider the case where we have two detector outputs, $x_1(t)$ and $x_2(t)$ ($t$ stands for time).
The former is the output from the laser interferometer, namely, the strain channel, 
and the latter is an environmental channel such as an output of a seismograph.
We wish to separate gravitational wave signal $h(t)$ and non-Gaussian noise
$k(t)$ using the data of $^t\vx(t)=(x_1(t),x_2(t))$.

As the simplest case we assume that there is a linear relation between the outputs
and the sources:
\beq
\vx(t)=\begin{pmatrix} x_1(t)\\ x_2(t)\end{pmatrix}=A\vs(t),~~~
\vs(t)=\begin{pmatrix} s_1(t)\\ s_2(t)\end{pmatrix}=\begin{pmatrix} h(t) + n(t)\\ k(t)\end{pmatrix},
  \label{vs}
\eeq
where $A$ is assumed to be a time independent matrix. Since the output of a laser interferometer, of course, suffers from Gaussian noise $n(t)$, we can regard $s_1(t) = h(t) + n(t)$ as an original signal. Note that non-Gaussian noise $k(t)$ can contain any Gaussian noise as a part of it. Thus, we have not added any Gaussian noise to $s_2(t)$ explicitly.

Since the gravitational wave is so weak that it will not affect any environmental meters such as a seismograph, one may set $A$ as
\beq
A=\begin{pmatrix} a_{11}& a_{12}\\ 0 & a_{22}\end{pmatrix}. \label{aform}
\eeq
The aim of ICA is to find a linear transformation
\beq
  \vy = W\vx, \label{eq.tr}
\eeq
such that two components of the transformed variables $\vy$
are statistically independent of each other. Here the distribution of $\vy$, $p_y(\vy)$, is constructed from the observed distribution function of $\vx$, $p_x(\vx)$, through the transformation~\eqref{eq.tr} as
\beq
p_y(\vy) \equiv ||W^{-1}||p_x(\vx),
\eeq
where $||X||$ denotes the determinant of matrix $X$.  
Thanks to the assumption (\ref{aform}), the matrix $W$ also takes a form
\beq
  W=\begin{pmatrix} w_{11}& w_{12}\\ 0 & w_{22}\end{pmatrix}. \label{wform}
\eeq
However, since we do not know all the component of $A$, we attempt to determine $W$ to be $A^{-1}$ in such a way that the components of $\vy$,
$y_1(t)$ and $y_2(t)$ to be statistically independent as much as
possible.
In~\cite{Morisaki:2016sxs} this was achieved by using the 
the Kullback-Leibler divergence~\cite{KL}, which represents a distance in the space of statistical distribution functionals. It is defined between two arbitrary PDFs, {\it e.g.}, $p_y(\vy)$ and $q(\vy)$ as 
\beq
 D[p_y(\vy); q(\vy)]=\int p_y(\vy)\ln \frac{p_y(\vy)}{q(\vy)}dy
=E_{p_y}\lkk \ln \frac{p_y(\vy)}{q(\vy)}\rkk.
\eeq
Here $E_{p_y}[\cdot]$ denotes an expectation value with respect to a PDF $p_y$.
Then we can obtain mutually independent variables $\vy$ by minimizing a cost function $L_q(W) \equiv D[p_y(\vy); q(\vy)]$, where $q(\vy) = q(y_1)q(y_2)$ is an appropriately chosen distribution function.

The most proper choice of $q(\vy)$ is obviously the true distribution function of independent source variables $\vs$, $r(\vs) = r_1[s_1(t)]r_2[s_2(t)]$, which is not known a priori. Because $n(t)$ is a Gaussian with vanishing mean in this simple setup, its statistical property is entirely characterized by the two-point correlation function $K(t-t')=\langle n(t)n(t')\rangle$. Then the marginal distribution function of $s_1(t)$ is given by
\beq
 r_1[s_1(t)]=\frac{1}{\sqrt{2\pi}\sigma}\exp\lkk -\frac{1}{2\sigma^2}
\lmk s_1(t)-h(t,\theta)\rmk^2\rkk,~~~\sigma^2=K(0), \label{gauss}
\eeq
where $h(t,\theta)$ is the actual waveform of gravitational radiation 
emitted from some source, where $\theta$ collectively denotes parameters of the source.
On the other hand, we do not specify the PDF of $k(t)$, $r_2(s_2)$, except that
it is a super-Gaussian distribution such as a Poisson distribution
with a larger tail than Gaussian. We show, however, that we can obtain the matrix $W$ easily for our particular problem with $a_{21}=w_{21}=0$ as we see below.

\subsection{Correlation method}\label{Corr} 
From now on we replace the ensemble 
average $E[\cdot]$ by temporal 
average of observed values of $\vx$ which we denote by brackets. For the true distribution $r(\vy)$, minimization of the cost function $L_r(W)$ results in decorrelating $y_1$ and $y_2$~\cite{Morisaki:2016sxs}, {\it i.e.} $\la y_1(t)y_2(t)\ra = 0$.

From 
\beq
\begin{pmatrix} y_1(t)\\ y_2(t)\end{pmatrix}=
\begin{pmatrix} w_{11}& w_{12}\\ 0 & w_{22}\end{pmatrix}
\begin{pmatrix} x_1(t)\\ x_2(t)\end{pmatrix}=
\begin{pmatrix} w_{11}x_1(t)+w_{12}x_2(t)\\ w_{22}x_2(t)\end{pmatrix}, \label{ymatrix}
\eeq
it is equivalent to 
requiring 
\beq
\langle y_1(t)x_2(t)\rangle 
=w_{11}\langle x_1(t)x_2(t)\rangle +w_{12}\langle x_2^2(t)\rangle =0.
\eeq
We therefore obtain
\beq
  w_{12}= -\frac{\la x_1x_2\ra}{\la x_2^2 \ra}w_{11}.  \label{1211}
\eeq
Since ICA does not uniquely determine the overall factor of $\vy$ by nature, this relation
suffices for our purpose to determine $y_1$.  These are what we calculated in our previous paper \cite{Morisaki:2016sxs}
using the Kullback-Leibler divergence.

Here we develop a multiple component method for further analysis and we apply it in \S~\ref{multi}. For three components, $\vy(t)$ and $\vx(t)$ become
\beq
\begin{pmatrix} y_1(t)\\ y_2(t)\\ y_3(t)\end{pmatrix}=
\begin{pmatrix} w_{11}& w_{12} &w_{13}\\ 0 & w_{22} &w_{23}\\0&w_{32}&w_{33}\end{pmatrix}
\begin{pmatrix} x_1(t)\\ x_2(t)\\x_3(t)\end{pmatrix}=
\begin{pmatrix} w_{11}x_1(t)+w_{12}x_2(t)+w_{13}x_3(t)\\ w_{22}x_2(t)+w_{23}x_3(t)\\w_{32}x_2(t)+w_{33}x_3(t)\end{pmatrix}, \label{ymatrix_3}
\eeq
and
\beq
\begin{pmatrix} x_1(t)\\ x_2(t)\\ x_3(t)\end{pmatrix}=
\begin{pmatrix} a_{11}& a_{12} &a_{13}\\ 0 & a_{22} &a_{23}\\0&a_{32}&a_{33}\end{pmatrix}
\begin{pmatrix} s_1(t)\\ s_2(t)\\s_3(t)\end{pmatrix}=
\begin{pmatrix} a_{11}s_1(t)+a_{12}s_2(t)+a_{13}s_3(t)\\ a_{22}s_2(t)+a_{23}s_3(t)\\a_{32}s_2(t)+a_{33}s_3(t)\end{pmatrix}. \label{xmatrix_3}
\eeq
In this case also, the minimization of cost function results in decorrelating $\vy$, $\la y_1y_2\ra = \la y_2y_3\ra = \la y_3y_1\ra  = 0$. This is achieved by the analogy of the Gram–Schmidt process which is a method for orthonormalising a set of vectors, and it can be extended to the case where there are more than three components.

Because of the gauge degree of freedom, we can take $w_{32} = 0$ without loss of generality and choose
\beq
y_3(t) = \tilde{x}_3(t) \equiv \frac{x_3(t)}{\sqrt{\la x_3^2\ra}}. 
\eeq
We first require $\la y_2(t)y_3(t) \ra = \la y_2(t)x_3(t) \ra = 0$. This gives following relation,
\beq
w_{23} = -\frac{\la x_2x_3\ra}{\la x_3^2\ra}w_{22}.
\eeq
Based on this, we can choose
\beq
y_2(t) = \tilde{x}_2(t) \equiv \frac{x'_2(t)}{\sqrt{\la x_2^{'2} \ra}},\ \ \ x'_2(t) \equiv x_2(t) - \frac{\la x_2x_3 \ra}{\la x^2_3\ra}x_3(t).
\eeq
 If we take
\beq
y_1(t) = x_1(t) - \la x_1\tilde{x}_2\ra \tilde{x}_2(t) -\la x_1\tilde{x}_3\ra\tilde{x}_3(t), \label{corr_3}
\eeq

$\la y_1(t)y_2(t) \ra= \la y_2(t)y_3(t)\ra = \la y_3(t)y_1(t) \ra = 0$ is satisfied. Note that Eq.\ (\ref{corr_3}) is symmetrical with respect to the permutation of $x_2(t)$ and $x_3(t)$.

Thus we can observe that the correlation method of ICA shown here is equivalent to the instantaneous Wiener filtering\footnote{The Wiener filtering adopted in~\cite{1999gr.qc.....9083A,2012RScI...83b4501D,2012CQGra..29u5008D,2014CQGra..31j5014M,2015CQGra..32p5014T} takes into account the time delay in transfer functions. We can easily incorporate it to our analysis, too, as already demonstrated in~\cite{Morisaki:2016sxs}.}, and this is due to the particular character of our problem that only the strain channel is sensitive to the gravitational wave signal with $a_{i1} = 0\ (i \neq 1)$ in our linear model.

\subsection{FastICA method}\label{FastICA}
Next, we introduce another method to obtain a matrix $W$ called FastICA \cite{FastICA}
which can be easily implemented even when $\vx(t) = A\vs(t)$ has more than two components. 
Note that this method can be applied to various cases of signal separation other than the case formulated in \S \ref{problem}.

In this method, assuming that each component, $s_i(t)$, of source vector $\vs(t)$ is
properly normalized with vanishing mean, 
we first apply whitening to the detector outputs $\vx(t)$
and take the dispersion of each source $s_i(t)$ to be unity. This is achieved in the following way.
First let the normalized eigenvector and corresponding 
eigenvalue of a matrix $\langle \vx\, ^t\vx\rangle$ be
${\vect c}_i$ and $\lambda_i$, respectively $(i=1,2,...)$, and define
a matrix $\Gamma$ by $\Gamma=({\vect c}_1, {\vect c}_2, {\vect c}_3,...)$, and $\Lambda^{-1/2}$ by $\Lambda^{-1/2}=\mathrm{diag}(\lambda_1^{-1/2}, \lambda_2^{-1/2},...)$.  Then the whitened variable $\tilde{\vx}(t)$
is defined by 
\beq
 \tilde{\vx}(t)=\Lambda^{-1/2} \ ^t\Gamma\vx
 =  \Lambda^{-1/2}\ ^t\Gamma A\vs \equiv \tilde{A}\vs,
 \eeq
which satisfies 
\beq
\langle\tilde{\vx}(t)\,^t\tilde{\vx}(t)\rangle 
=\langle \tilde{A}\vs\,^t(\tilde{A}\vs)\rangle =\tilde{A}\langle \vs\, ^t\vs\rangle ^t\tilde{A}=
\tilde{A}\ ^t\tilde{A}=E.
\eeq
Here we have used the statistical independence of each component of the normalized
source term $s_i$.
This means that the matrix $\tilde{A}$ is an orthogonal matrix
and that $W$ may be identified with $ ^t\tilde{A}$ for whitened 
output data $\tilde{\vx}$.  Thus we may restrict $W$ to be an orthogonal
matrix, too, after appropriate whitening\footnote{Note that this procedure is also called as sphering and has nothing to do with the whitening of strain data in frequency domain.}.

We here choose $q(\vy)$ as a product of marginal distributions,
\beq
 q(\vy)= \tilde{p}_y(\vy)\equiv \prod_i \tilde{p}_i(y_i),~~~
 \tilde{p}_i(y_i)=\int p_y(\vy)dy_1...dy_{i-1}dy_{i+1}...,  
 \eeq
since $p_y(\vy) = \tilde{p}_y(\vy)$ is the condition for statistical independence of the variables $\vy$. Then, the cost function defined in terms of the Kullback-Leibler divergence reads
\beq
 L_{\tilde{p}}(W)=D[p_y(\vy);\tilde{p}_y(\vy)]=-H[\vx]-\ln||W||+\sum_i H_i[y_i],
\eeq
where $H[\vx] \equiv -\int d\vx p_x(\vx)\ln p_x(\vx)$ is the entropy of the distribution of $\vx$, and $H_i[y_i]\equiv -\int dy_i \tilde{p}_i(y_i)\ln \tilde{p}_i(y_i)$ is the entropy 
of the marginal distribution of  $y_i$.
When $W$ is an orthonormal matrix, only the last term matters to determine $W$.
Hence minimization of the cost function for $W$ is achieved by minimizing
 entropy of the marginal distribution of each variable.  This is the spirit of
 the FastICA  method.  
 It has been
proposed to maximize the negentropy defined by
\beq
 J[y_i] \equiv H[\nu]-H[y_i], 
 \eeq
 which is a positive semi-definite quantity, instead of the entropy itself.
 Here $\nu$ is a random Gaussian variable with vanishing mean and unit variance.

 In order to achieve easier implementation of the method, however,
 we minimize a simpler cost function $L(\vw_i)$ for each row vector $\vw_i$ constituting
 the matrix $W$ as $W\equiv (\vw_1,\vw_2,...)$.  Since $W$ is an orthogonal matrix
 now, we find $|\vw_i|^2=1$, so the cost function may be defined as
 \beq
  L(\vw_i)=\lnk E[G(y_i)]-E[G(\nu)] \rnk^2-\beta \lkk |\vw_i|^2-1 \rkk,  \label{costw}
  \eeq
  where $G$ is an appropriate nonquadratic function and  $\beta$ is a Lagrange multiplier.
 Minimization of Eq. (\ref{costw}) corresponds to solving the following equation:
 \beq
  E[\tilde{\vx}g(^t\vw_i\tilde{\vx})]-\beta \vw_i=0,
  \eeq
where $g(y)=G\rq{}(y)$.  FastICA solves for this equation starting from an arbitrary initial choice of $\vw_i$ in terms of the Newton method.
  

 \section{Analysis of iKAGRA data}\label{result} 
\begin{figure}[H]
\centering
 \subfloat[Strongly correlated data]{\includegraphics[width=8.3cm]{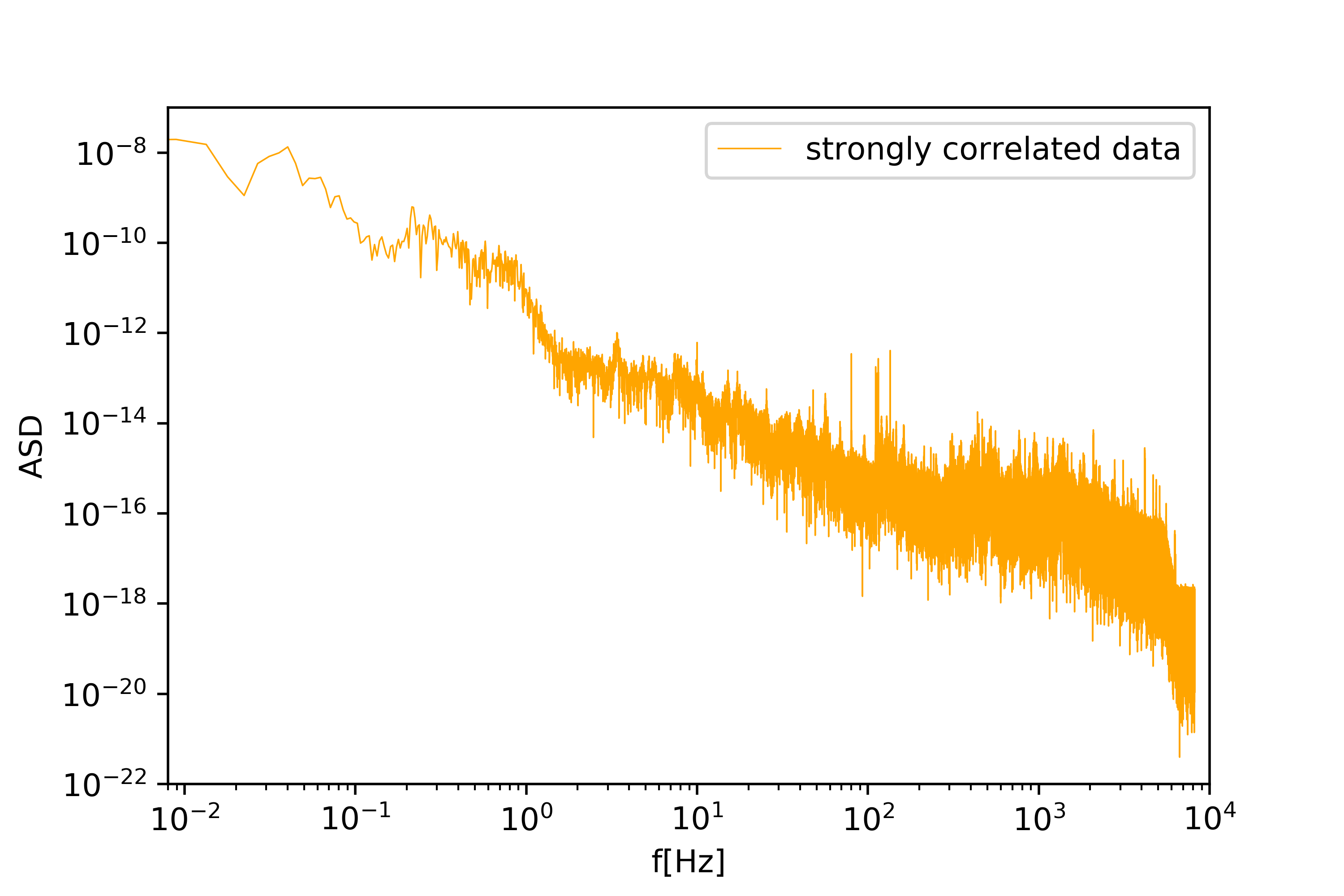}}~
 \subfloat[Weakly correlated data]{\includegraphics[width=8.3cm]{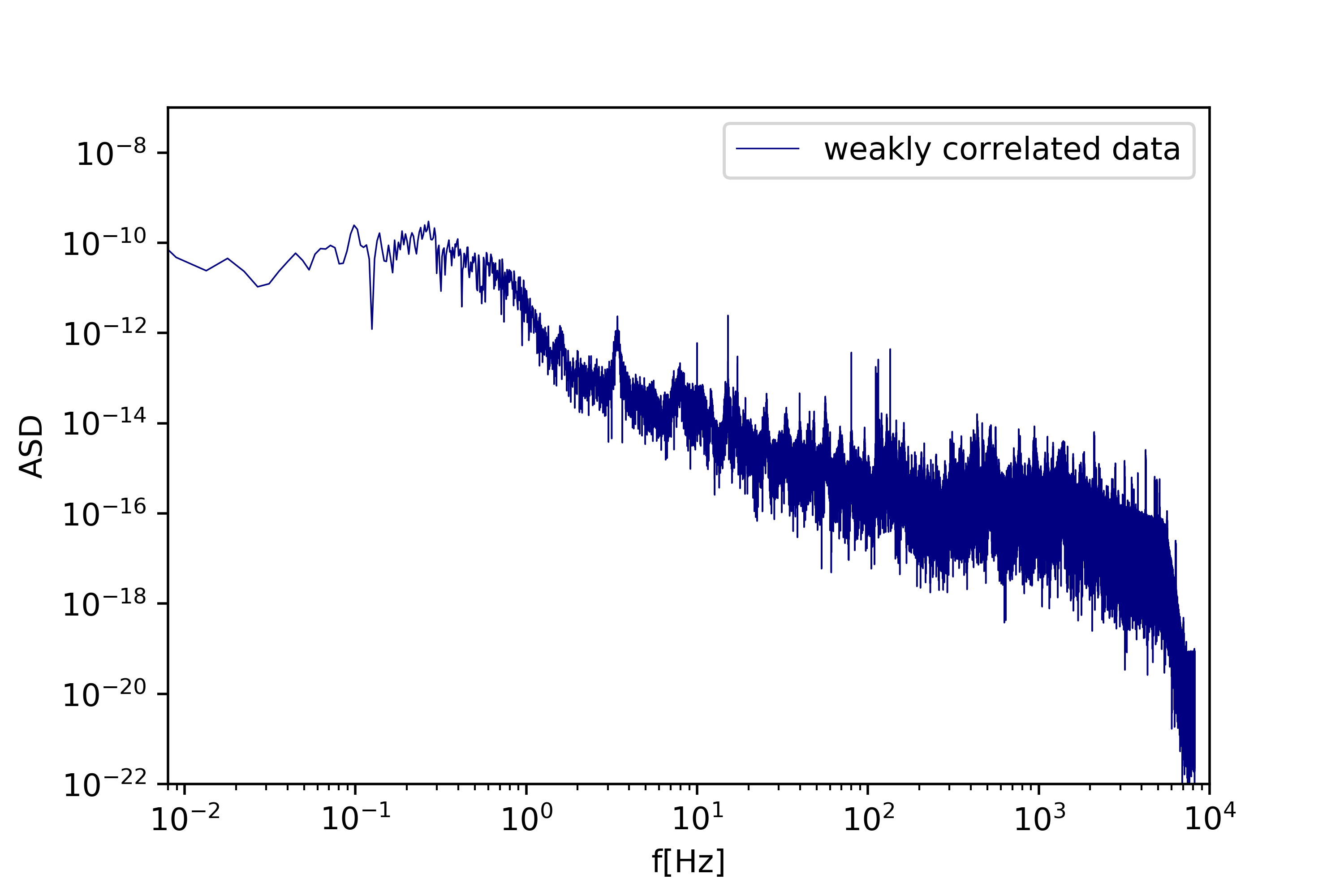}}
 \caption{ASDs of strain channels for two datasets. For the strongly correlated data, ASD below 0.1Hz becomes much larger than that of the weakly correlated data. This means that the strongly correlated data is contaminated by seismic noise at lower frequencies.}\label{fig10}
\end{figure}

The initial engineering run of KAGRA without the cryogenic system
was done in March and April, 2016 \cite{Akutsu:2017kpk}.
From the results of many time series data that we analyzed, we report those of two datasets of 224 second long. One starts from 20:15:11 UTC on April 14, 2016. The other starts from 01:01:35 UTC on April 17, 2016. For each dataset, we calculated Pearson's correlation between the strain channel and each of 35 physical environmental monitor (PEM) channels. We found that almost all these channels in the latter (former) data set strongly (weakly) correlated with the strain channel. We call the latter (former) the strongly (weakly) correlated data
. The amplitude spectrum density (ASD) of the strain channel for each data set is depicted in Fig.\ \ref{fig10}.

We chose two channels which showed large correlation with the strain channel for each dataset. Those channels are listed in Table 1.
\begin{table}[H]
  \centering
  \caption{Correlation between PEM channels and strain.}
  \begin{tabular}{|c|c||c|}\hline
    dataset & channel & correlation coefficient \\ \hline \hline
    strongly  & {\tt PEM-EX\_SEIS\_Z\_SENSINF\_OUT16} (4724ch)& $-0.6409$\\ \cline{2-3}
    correlated& {\tt PEM-EY\_SEIS\_WE\_SENSINF\_OUT16} (4823ch) & $0.5892$\\ \hline \hline
    weakly  & {\tt PEM-EX\_SEIS\_Z\_SENSINF\_OUT16} (4724ch) & $0.3078$\\ \cline{2-3}
    correlated& {\tt PEM-EY\_SEIS\_NS\_SENSINF\_OUT16} (4774ch)& $-0.2312$ \\ \hline
  \end{tabular}
\end{table}
For both datasets, 4724ch had the largest correlation with the strain. This channel is the output of the seismograph that observes vertical vibration installed at the end of the X arm. Both 4774ch and 4823ch are the outputs of the seismographs installed at the end of the Y arm, and they observe horizontal vibration orthogonal to each other.

We have made mock strain data injecting sinusoidal continuous waves
\begin{equation}
  s(t) = A\sin(2\pi ft),
\label{sin}
\end{equation}
to the strain channel and applied two methods of ICA, which were introduced in the previous section, to this mock data and those environmental channels.

We utilized the python implementation of FastICA from {\tt scikit-learn}\footnote{https://scikit-learn.org/stable/}. We found that results often depend on initial conditions where the Newton method is started. To mitigate this, we parallelly generated at most thirty realizations and chose one which gives the highest SNR.

\subsection{Global performance}\label{global}
\begin{figure}[H]
  \centering
  \includegraphics[width=13cm,clip]{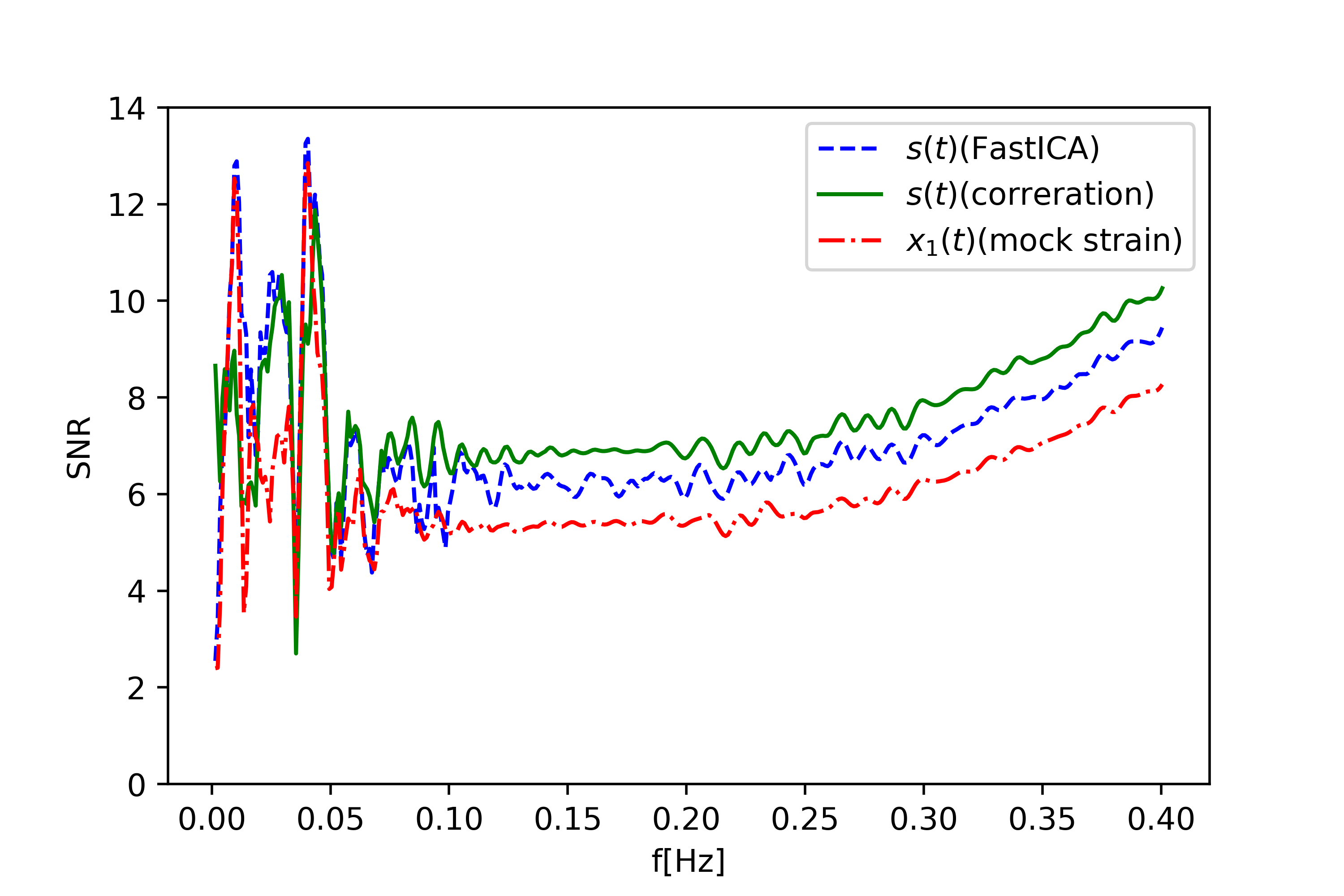}
    \caption{SNR for varying $f$ with and without ICA using 4724ch for the strongly correlated dataset.
   The red line corresponds to the raw mock strain, while the green and
   blue lines are noise-removed data using the correlation method and
   FastICA, respectively.}
    \label{fig1}
\end{figure}

First, we analyze how the signal-to-noise ratio (SNR) changes before and after noise separation by ICA for mock data with varying frequencies $f$. We performed matched filter (MF) analysis to both the raw mock strain data
and the noise-removed data in terms of the two methods of ICA using 4724ch as an environmental channel. For various $f$ of injected signal~\eqref{sin}, we calculated SNR by applying MF with the same frequency as the injected signal. We simultaneously plot the results against the data before and after ICA to assess the global performance of ICA. For strongly correlated data, the results are shown in Fig.~\ref{fig1}.

In this dataset, strain had larger amplitude than the other dataset, and we set $A = 9 \times 10^{-10}$. As one can see from Fig.\ \ref{fig1}, SNRs are homogeneously enhanced by ICA for $f \gtrsim 0.1$Hz. The correlation method enhances the SNR more than FastICA.
However, there are anomalous peaks at frequencies 0.01Hz and 0.04Hz. As shown in Fig.\ \ref{fig10}(a), even in the absence of injection the strain channel has large amplitude at these frequencies, which is predominantly contributed by seismic noises. We also found that their oscillation phases are more or less stable during the time period we analyzed. Such noises are difficult to be distinguished from our sinusoidal signal waveform and hence yield large SNR of mock strain as shown in Fig.\ \ref{fig1}. This, however, indicates that by removing contribution of the noises, the SNR can possibly be reduced rather than enhanced provided the injected signal is moderate. This is actually realized in the analysis based on the correlation method as seen in Fig.\ \ref{fig1}.

On the other hand, in the case of FastICA, the reduction of SNR is not seen. This is solely due to our implementation, which tries to increase the SNR as much as possible as mentioned before. In that sense, around the 0.01Hz and 0.04Hz peaks, blue line in Fig.\ \ref{fig1} corresponds to the SNR of the separated noise.

Apart from these low frequencies contaminated by seismic noises, we find that ICA improves SNR significantly throughout the entire frequency range with $f \gtrsim 0.1$Hz. However, based on these considerations, it is deduced that ICA works even near the peak due to seismic noise.

\begin{figure}[H]
  \centering
    \includegraphics[width=13cm,clip]{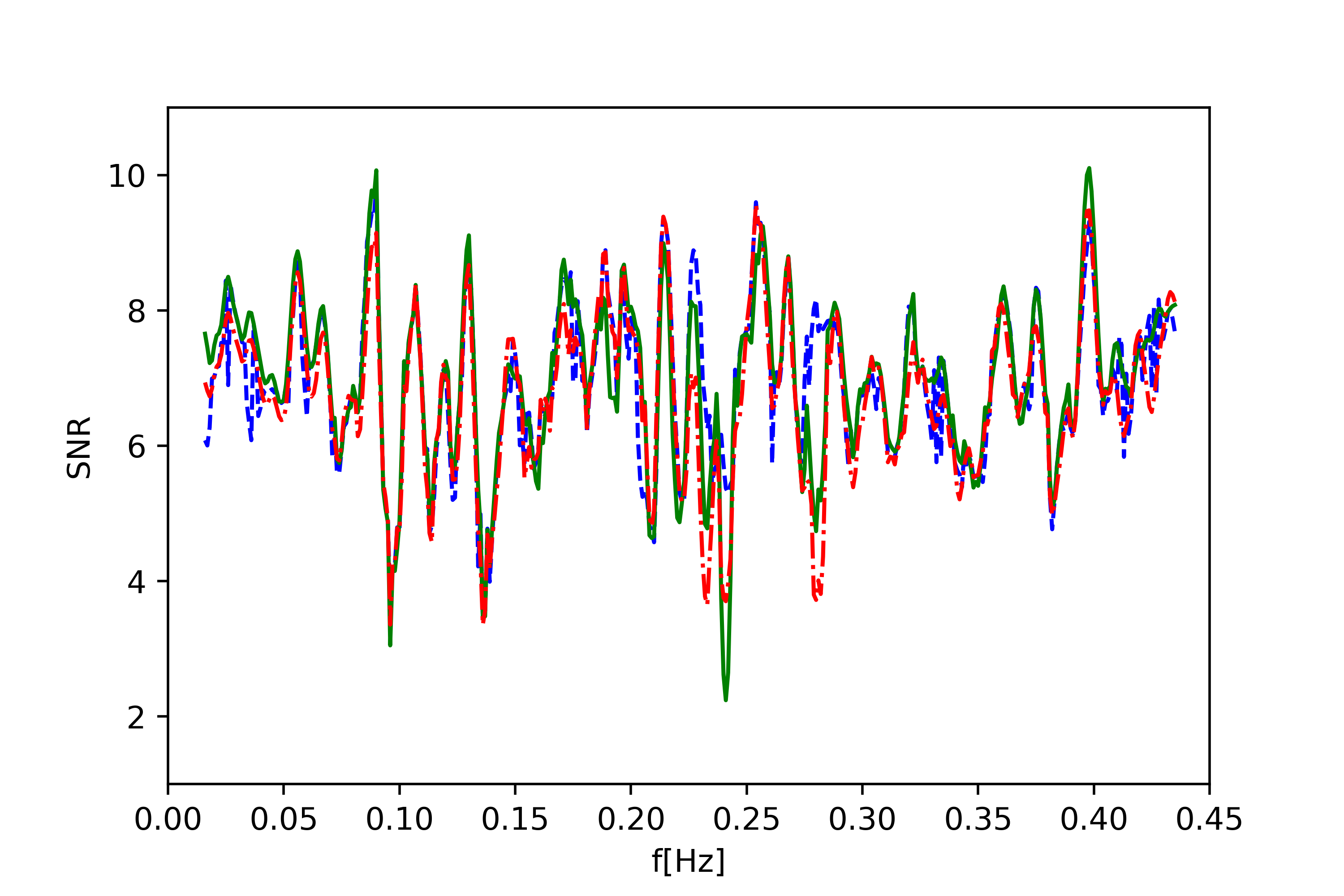}
    \caption{The same figure as in Fig.\ \ref{fig1} but for the weakly correlated dataset.}
    \label{fig2}
\end{figure}

For the weakly correlated data, the results are shown in Fig.\ \ref{fig2}. The amplitude of strain at this time period is moderate, and we set $A = 3 \times 10^{-11}$. As is seen in Fig.\ \ref{fig2}, the SNR of the data with ICA is higher than the mock data in several frequency ranges. Comparing FastICA with the correlation method, the correlation method has fewer frequencies where the SNR falls below that of mock data.
\begin{figure}[H]
  \centering
    \includegraphics[width=13cm,clip]{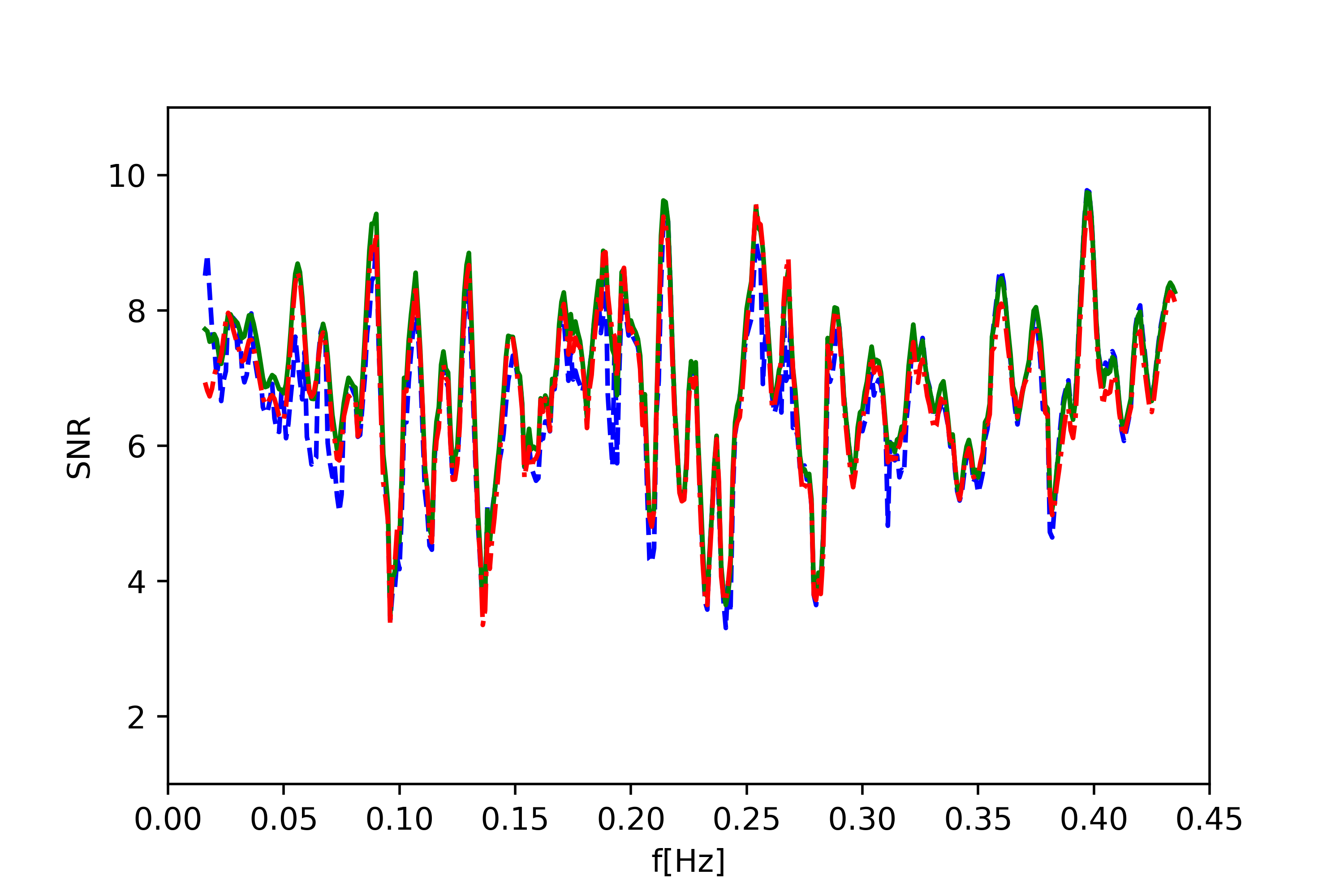}
    \caption{Same as in Fig.\ \ref{fig2} but using 4774ch.}
    \label{fig3}
\end{figure}
As for the weakly correlated data, 4774ch had the second highest correlation with strain. If we use 4774ch instead of 4724ch as the environmental data, the result changes as shown in Fig.\ \ref{fig3}. Compared with the case 4724ch is used (Fig.\ \ref{fig2}), the frequency region where the SNR rises is different. As a whole the improvement of SNR is less significant, which is a natural result considering that the correlation coefficient of 4774ch is smaller than that of 4724ch.

\subsection{Parameter estimation for strongly correlated data}\label{strong}
\subsubsection{Two channels ICA}\label{two}
Next, we perform parameter estimation using the strongly correlated data to examine whether ICA can recover correct parameters of injected signals. We injected the sinusoidal waveform in Eq. (\ref{sin}) with $f = 0.125$Hz and $A = 1.3\times10^{-9}$. 
We applied MF analysis to search for the
frequency with the highest SNR which corresponds to the maximum likelihood estimation of the parameter.
We compare how the result of parameter estimation changes before and after ICA and how much the SNR changes.

\begin{figure}[H]
  \centering
    \includegraphics[width=12cm,clip]{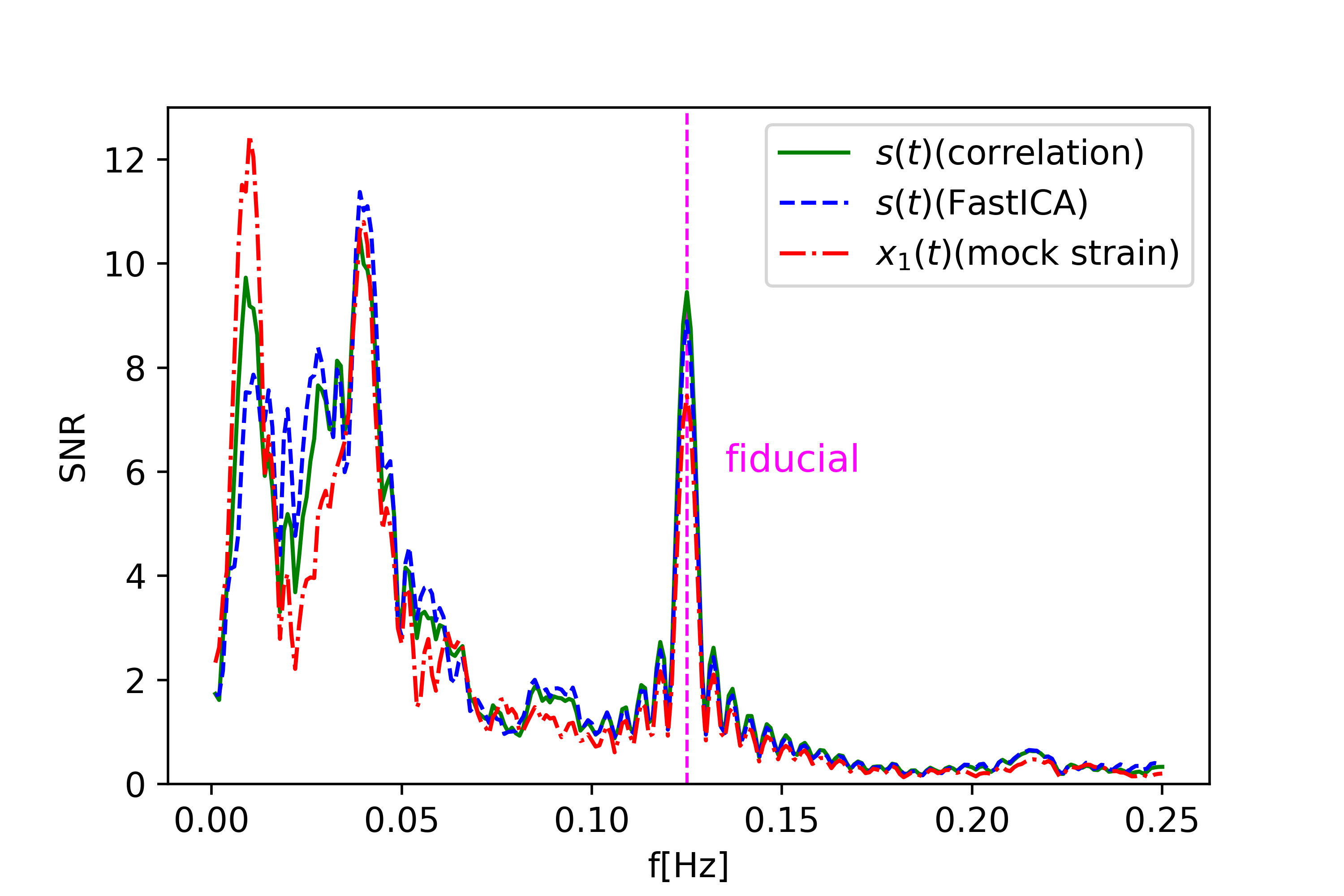}
    \caption{Parameter estimation with fiducial frequency f=0.125Hz. Correspondence of each line is the same as in Fig.\ \ref{fig1}.}
    \label{fig4}
\end{figure}

Figure\ \ref{fig4} depicts the SNR before and after applying ICA.

In this case, we can see the effect of seismic noise directly. By ICA with 4724ch, SNR at $f \sim 0.01$Hz is reduced and that at the injected frequency $f = 0.125$Hz is successfully enhanced. From this result, we deduce that 4724ch is highly correlated to the 0.01Hz peak. On the other hand, the peak of $0.04$Hz is still higher, which turned out to be correlated to 4823ch which had the second largest correlation with the strain, as we will see below.

\subsubsection{Multiple channels ICA}\label{multi}
\begin{figure}[H]
  \centering
    \includegraphics[width=12cm,clip]{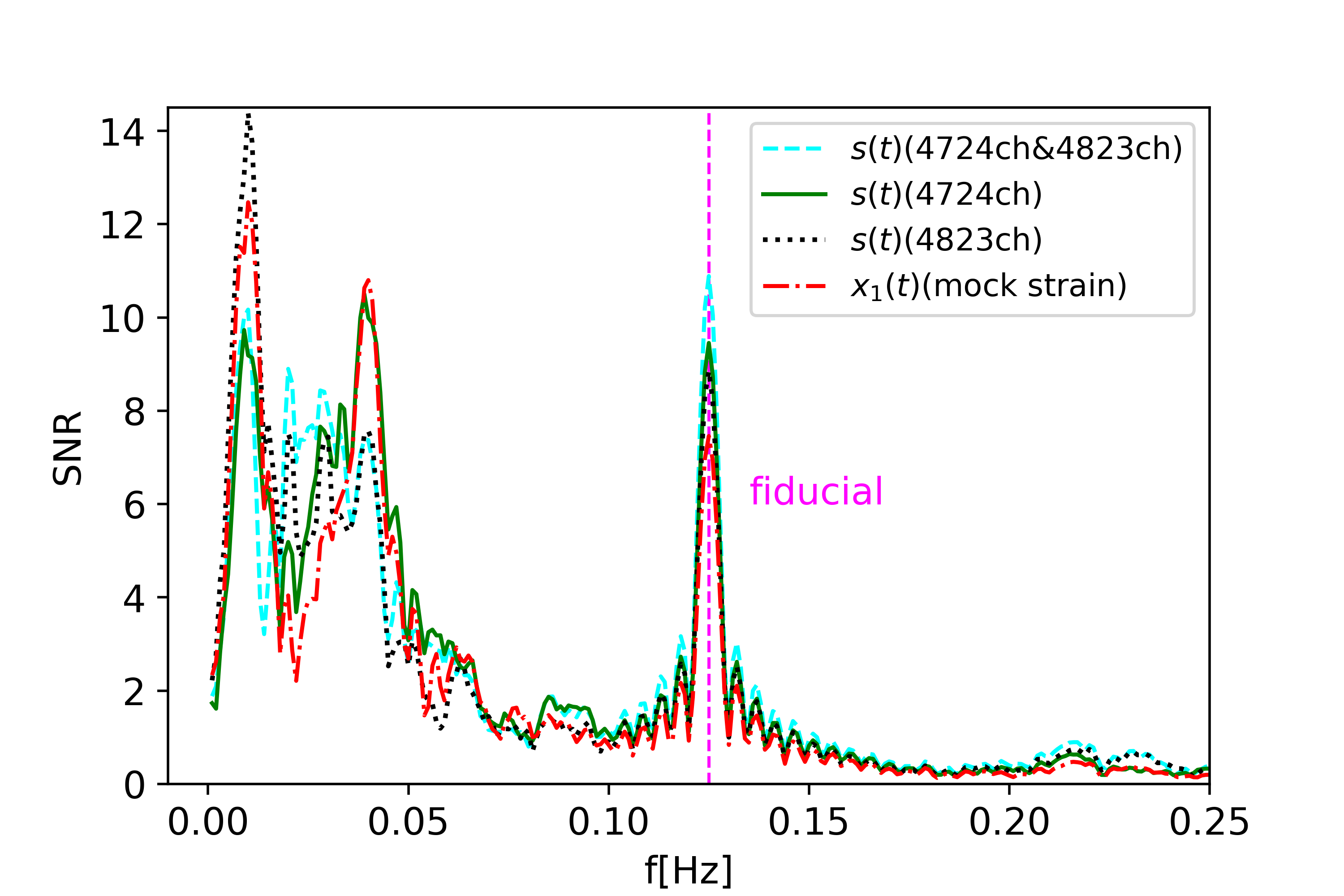}
    \caption{Parameter estimation with multiple channels ICA(correlation method).}
    \label{fig5}
\end{figure}

\begin{figure}[H]
  \centering
    \includegraphics[width=12cm,clip]{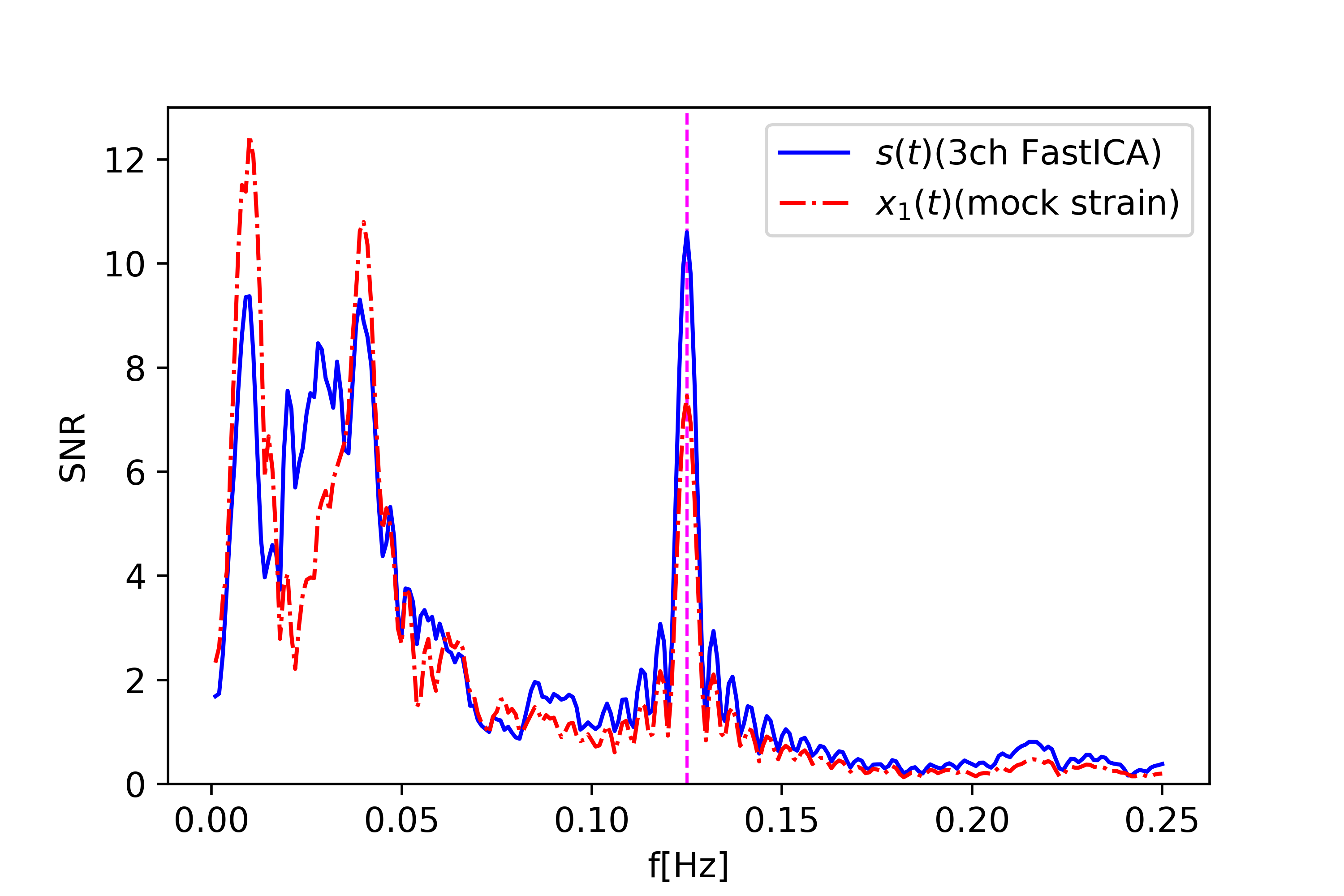}
    \caption{Parameter estimation with multiple channels ICA(FastICA).}
    \label{fig12}
\end{figure}

・\textbf{Correlation method}\\
As is seen \S~\ref{global}, the correlation method shows more
stable performance than FastICA, although it is much simpler.
This method can be generalized to multi-channel analysis. As a first step to multi-channel analysis, here we investigate the effectiveness of three components analysis, which we developed in \S \ref{Corr}, including two PEM channels which strongly correlated to the strain. 
For this purpose we have used the mock data including the same signal waveform as in the previous subsection, and applied the three components correlation method to this mock data, 4724ch and 4823ch. 
The result is depicted in Fig.\ \ref{fig5}. We simultaneously plotted the results of two-component analysis in which we used 4724ch and 4823ch respectively. 
The green and black lines correspond to the cases where noises are removed using one PEM channel. 
While the 0.01Hz peak was reduced by using 4724ch, the 0.04Hz peak was reduced by using 4823ch. However, both peaks cannot be reduced when we use only one PEM channel.
The data with ICA using two PEM channels (cyan line) has much higher SNR than the data with ICA using only one PEM channel. In addition, we successfully reduced both 0.01Hz peak and 0.04Hz peak. This result suggests that by combining many environmental channels we can effectively remove noises with various characteristic frequencies.\\
\\
・\textbf{FastICA}\\
As explained in \S~\ref{FastICA}, FastICA can be easily implemented even when there are more than two components. We applied FastICA to the mock data, 4724ch and 4823ch simultaneously. Here, mock data included the same sinusoidal signal as in the previous section. The result is shown in Fig.\ \ref{fig12}.

As compared to Fig.\ \ref{fig4}, SNR at fiducial frequency is much higher than the case where only 4724ch was used. In addition, its value is close to the that for three component correlation method (10.60 for FastICA, 10.89 for the correlation method). This result suggests that the use of multiple environmental channels can also enhance the effect of FastICA noise separation. However, compared to the case of the three components correlation method, we may have to make several trials of 3ch FastICA to obtain the best result. This indicates that correlation method is more effective than FastICA for this dataset.

\subsection{Parameter estimation for weakly correlated data}\label{weak}
\begin{figure}[H]
  \centering
    \subfloat[Overall view of the result.]{\includegraphics[width=12cm]{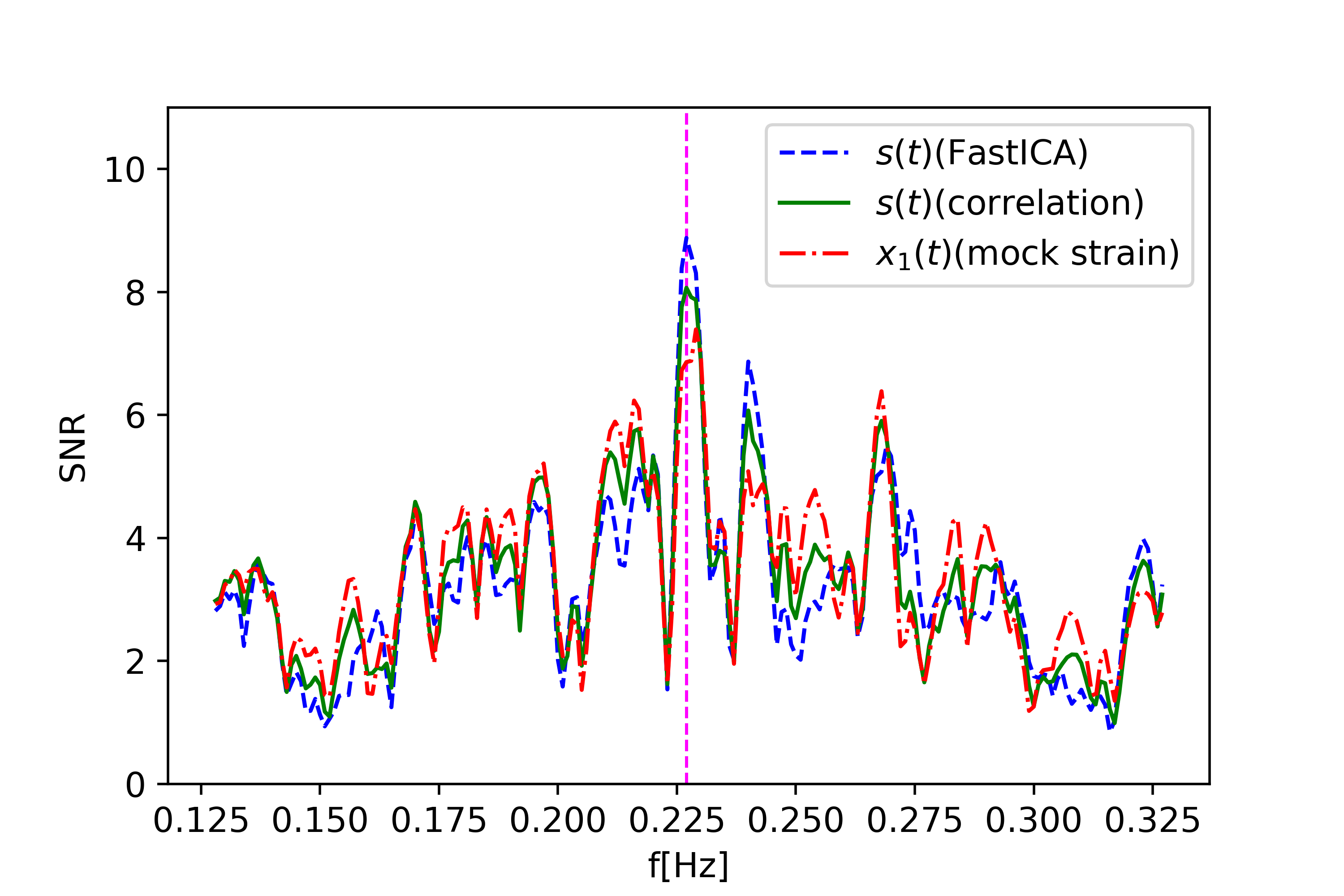}}\\
 \subfloat[Around the fiducial frequency.]{\includegraphics[width=12cm]{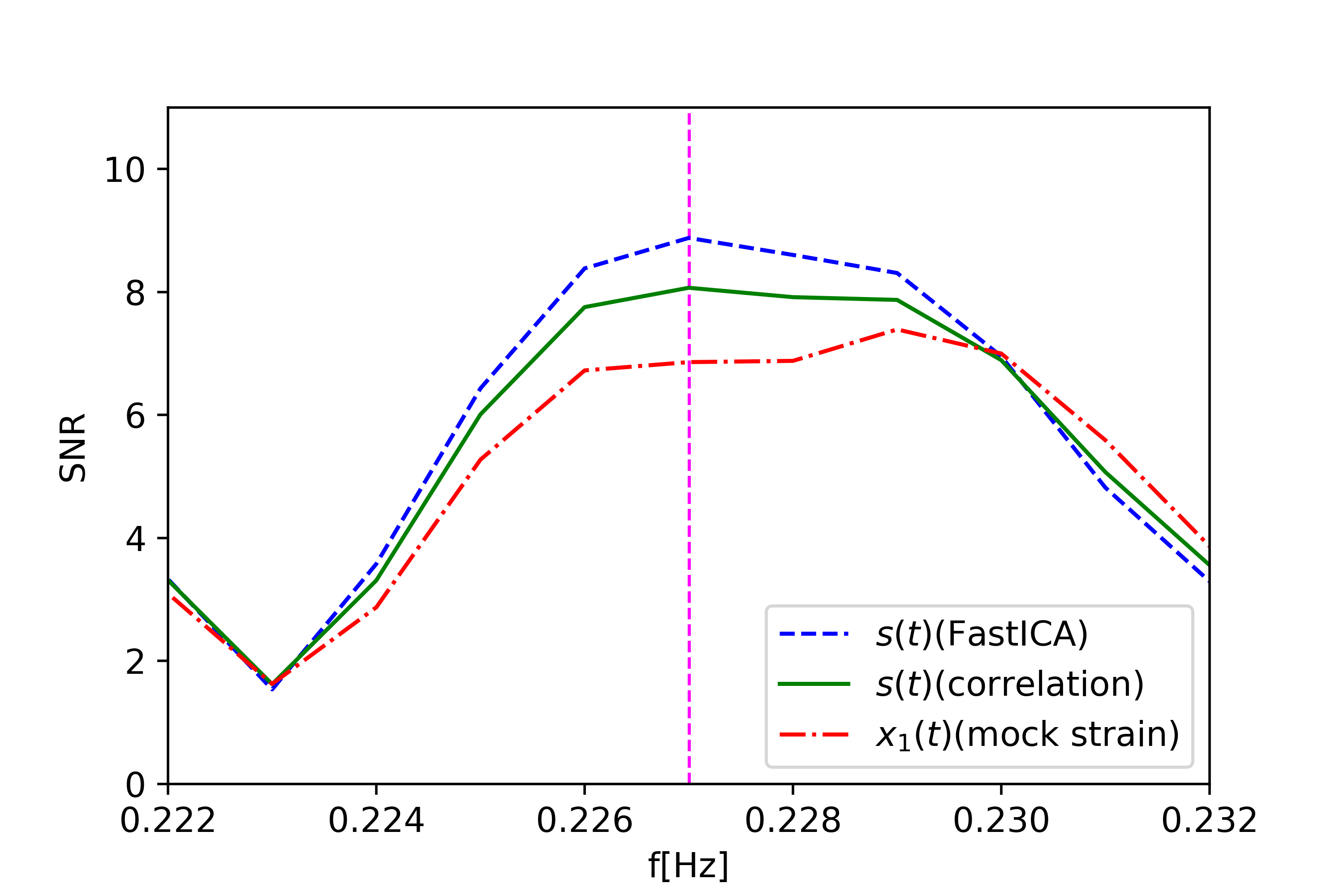}}
 \caption{Parameter estimation for the weakly correlated data with the fiducial frequency f=0.227Hz.}
 \label{fig6}
\end{figure}

We also perform parameter estimation for weakly correlated data. Here, we used 4724ch as an environmental channel. From Fig.\ref{fig2}, ICA using 4724ch is most effective for $f = 0.227$Hz with this dataset. We injected sinusoidal wave signal with $f = 0.227$Hz and $A = 3 \times 10^{-11}$. Again, we applied MF to search for the frequency with the highest SNR. The result is depicted in Fig.\ \ref{fig6}.\\
　The red line represents SNR calculated with the raw mock strain. The green and blue lines correspond to the noise-removed strain by the correlation method and FastICA, respectively. An enlarged figure of the fiducial ($f = 0.227$Hz) area is shown in Fig.\ \ref{fig6} (b). 

\begin{figure}[H]
  \centering
    \subfloat[Overall view of the result.]{\includegraphics[width=12cm]{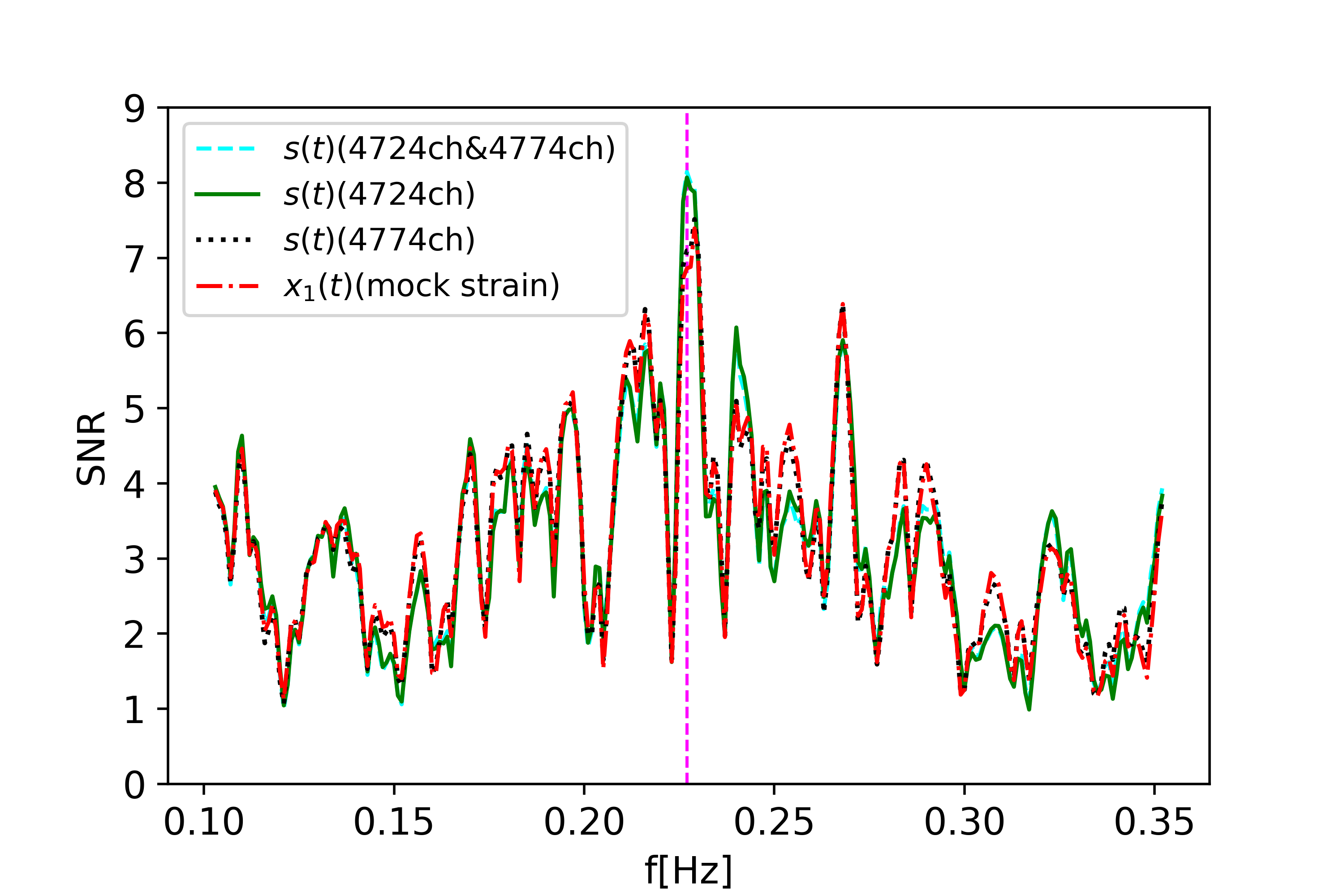}}\\
 \subfloat[Around the fiducial frequency.]{\includegraphics[width=12cm]{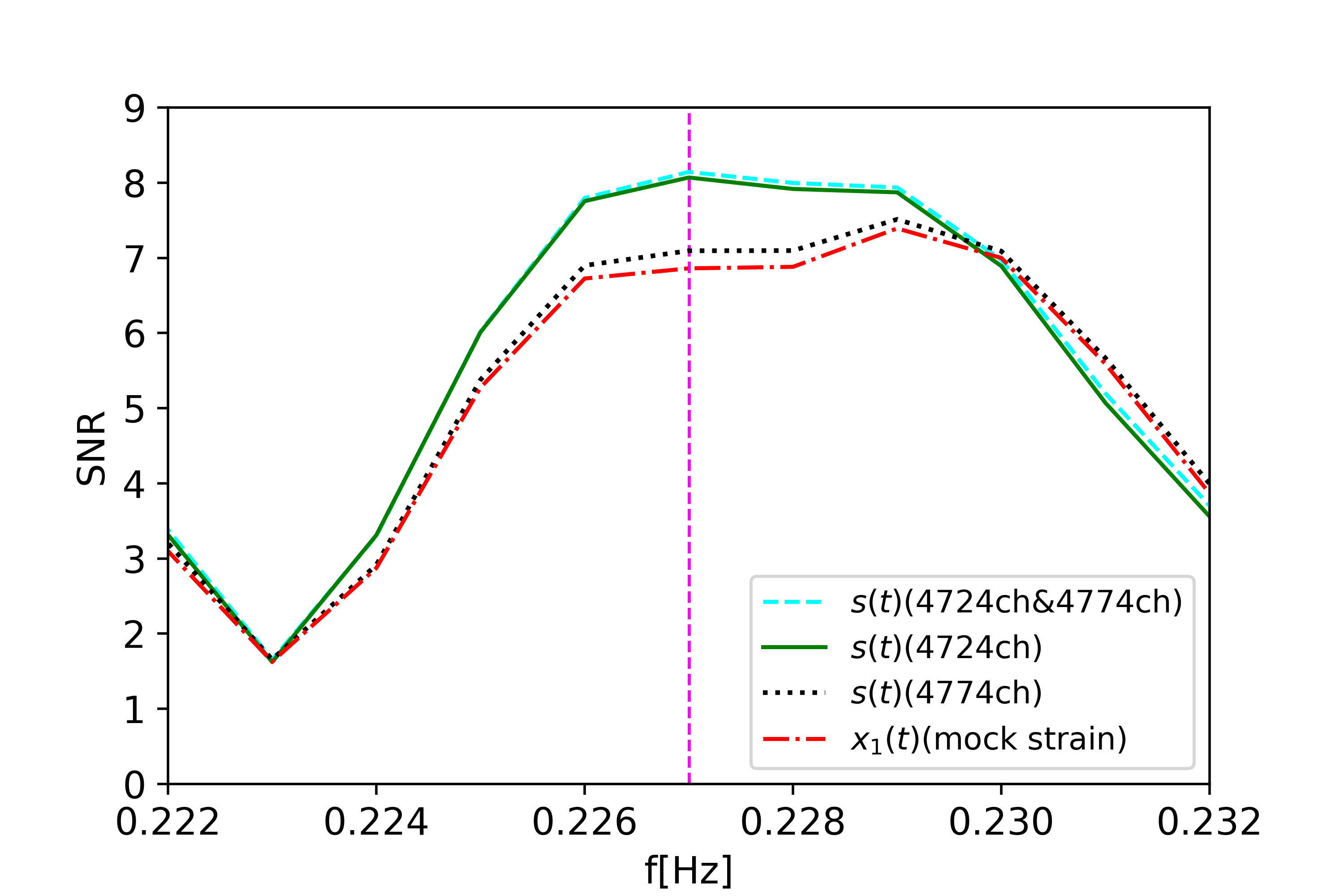}}
 \caption{Parameter estimation with the multiple-channel ICA(correlation method)}
 \label{fig8}
\end{figure}

As one can see, in the case of the raw mock strain, the position of the SNR peak deviates from the fiducial one. 
On the other hand, after applying ICA, the SNR is increased and the peak is found at the correct frequency.\\
　Next, we applied multiple channels ICA for this data using 4724ch and 4774ch. Here, we used the correlation method. Figure\ \ref{fig8} depicts the results of analysis. The enlarged figure of the fiducial area is shown in Fig.\ \ref{fig8} (b). 

The green and black lines correspond to the data with ICA using one PEM channel. When using only 4774ch, enhancement of SNR is small and still the SNR peak deviates from the fiducial frequency. However, the data with ICA using two channels has slightly higher SNR at the correct frequency than the data with ICA using only 4724ch. This result for the weakly correlated data also supports our expectation that the effect of ICA can be enhanced by combining many environmental channels. 

\section{Discussion}\label{discuss}
In \S~\ref{result}, we show the performance of ICA as a method of non-Gaussian noise subtraction in GW data. Both ICA methods, namely the correlation method and FastICA subtract the portion of seismic noise. However, the correlation method shows better performance than that of FastICA in most cases. In this section, we consider the reason why this difference appears.

We here use the same notation as in \S~\ref{ICA}, $x_1(t)$ to be the strain channel, $x_i(t)\ (i = 2, ..., n)$ to be the other environmental channels. As we already discussed in \S~\ref{problem}, the data of these channels can be written in the following form
\beq
\begin{split}
  x_1(t) &= h(t) + n(t) + \sum_{j=2}^na_{1j}s_j(t),\\
  x_i(t) &= \sum_{j = 2}^na_{ij}s_j(t).
\end{split}
\eeq
Here $s_i(t)\ (i = 2, ..., n)$ are environmental noises that can be measured by the PEM channels $x_i(t)$, and $n(t)$ collectively represents noises of the strain channel to which these PEM channels are insensitive. 
Let us transform $x_1$ as
\beq
\tilde{x}_1 = x_1 + \sum_{j = 2}^nb_{1j}x_j, \label{non_corr}
\eeq
in order to satisfy $\la \tilde{x}_1x_i\ra = 0\ (i = 2, ..., n)$. This condition can be expanded as
\beq
\begin{split}
  \la\tilde{x}_1(t)x_i(t)\ra &= \left\la\lmk h(t) + \sum_{j=2}^na_{1j}s_j(t) + \sum_{j = 2}^nb_{1j}\sum_{l = 2}^na_{jl}s_l(t)\rmk \sum_{k = 2}^na_{ik}s_k(t)\right\ra\\
  &= \sum_{j=2}^na_{1j}a_{ij} + \sum_{j=2}^nb_{1j}\sum_{k=2}^na_{jk}a_{ik} = 0,
\end{split}\label{}
\eeq
where we have used $\la s_is_j\ra = \delta_{ij}$ as in \S~\ref{FastICA}. From this equation, we obtain
\beq
b_{1j} = \sum_{i = 2}^na_{1i}a_{ij}^{-1}
\eeq
with $j = 2, ..., n$.
Note that $a_{ij}^{-1}$ is the inverse matrix of $a_{ij}\ (i,j = 2, ..., n)$, which is an $(n-1)\times(n-1)$ partial matrix of the mixing matrix $A = (a_{ij})_{1 \leq i,j \leq n}$. By substituting this into eq.\eqref{non_corr}, we obtain
\beq
\begin{split}
  \tilde{x}_1 &= x_1 + \sum_{i = 2}^nb_{1i}x_i\\
  &= h(t) + n(t) + \sum_{i = 2}^na_{1i}s_i(t) + \sum_{i = 2}^n\sum_{j = 2}^n\sum_{l=2}^na_{1j}a_{ji}^{-1}a_{il}s_l(t)\\
  &= h(t) + n(t) + \sum_{i=2}^na_{1i}s_i(t) - \sum_{j = 2}^n\sum_{l=2}^na_{1j}\delta_{jl}s_l(t)\\
  &= h(t) + n(t) + \sum_{i=2}^na_{1i}s_i(t) - \sum_{j = 2}^na_{1j}s_j(t) = h(t) + n(t),
\end{split}
\eeq
which shows that all environmental noises $s_i$, which are measurable by PEM channels, are removed from $\tilde{x}_1$ just by imposing $\la \tilde{x}_1x_i\ra = 0\ (i = 2, ..., n)$. In other words, when we consider auxilialy channels which are not sensitive to gravitational waves, and their target noises affects the strain linearly and additively, we can obtain independent component $h(t)$ by the transformation~\eqref{non_corr} which eliminates the two-point correlation between the strain and those channels. Although we have not given concrete expression of the transformation~\eqref{non_corr}, $\la\tilde{x}_1x_i\ra = 0$ is naturally achieved by the correlation method which is analogous to the Gram-Schmidt orthogonalization. Thus, we find that the correlation method is the optimal filter for linearly coupled noise with $a_{i1} = 0$.

On the other hand, FastICA maximizes negentropy after the whitening which makes $\la\tilde{x}_1x_i\ra = 0$ without using the condition $a_{i1} = 0\ (i \neq 1)$. Since we do not recover this property in general even after maximizing the negentropy, FastICA tends to show less enhancement of SNR than that of the optimal correlation method for most cases in our analysis. This illustrates the importance of incorporating characteristic features of the system as much as possible before applying ICA.

However, the above discussion is only the case where linearly coupled noise with $a_{i1} = 0$ is concerned. As for real observational data, there should be much more complicated mixing such as nonlinear coupling of the noise, and we might need a formulation of ICA which treats general mixing of signals. In that sense, it is noteworthy that FastICA, which is formulated without any assumption like $a_{i1} = 0$, also shows enhancement of the SNR and improvement of the performance with multi-environmental channels to some extent.


\section{Conclusion}\label{conclusion}
In the present paper, we have demonstrated usefulness of ICA in gravitational wave  data analysis in application to the iKAGRA strain and environmental channels. Assuming continuous waves as input signals, we have shown that ICA can enhance SNR in particular when the strain channel has large correlation with environmental ones.
Moreover, we have shown that ICA can correctly recover input frequencies in parameter estimation. We have also found that combining multiple environmental channels can enhance the effect of ICA to improve SNR.
 
There are, however, a number of limitations in the analysis presented here because the iKAGRA data contains more low 
frequency mode than wanted due to the simplified vibration isolation system compared with the full designed specification which will be realized with bKAGRA~\cite{bKAGRA}, and iKAGRA configuration was not equipped with environmental monitors that measure hecto-Hertz frequencies.
Hence we had to concentrate on relatively lower frequency components as the first step of application of ICA to real 
data analysis of laser interferometers.

Another limitation is that we have restricted to the case all the environmental noises that can be measured by the PEM channels under consideration act on the strain channel linearly and additively, without incorporating nonlinear couplings. In this particular situation, we have shown that the Gram-Schmidt decorrelation approach, or the instantaneous Wiener filtering which we dubbed the correlation method, gives the optimal result of environmental noise removal as an implementation of ICA. However, ICA can be used even in the cases noises of different origin are nonlinearly coupled to affect the strain channel as demonstrated in \cite{Morisaki:2016sxs}.  This is one of the merits of
ICA absent in other methods.  We could not perform such an analysis here due to the limitation of available PEM channels.
We plan to return to this issue when the full cryogenic configuration of bKAGRA starts operation with more PEM channels. 
\if
In this paper we analyzed low frequency range mainly because iKAGRA measured mostly the low frequency seismic noises due to the simplified vibration isolation system compared with the full designed specification which will be realized with bKAGRA. Another reason is that only PEM channels measuring low frequency part were available for iKAGRA. These two problems will be solved in bKAGRA with improved vibration isolation system and various PEM channels, and we should be able to apply ICA in the hectoHerz frequency range relevant to GW physics. We will analyze O3 KAGRA data and apply ICA to them.
\fi

\vskip 1cm
\noindent
{\large\bf Acknowledgements}\\
This work was supported by
MEXT, 
JSPS Leading-edge Research Infrastructure Program, 
JSPS Grant-in-Aid for Specially Promoted Research 26000005, 
JSPS Grant-in-Aid for Scientific Research on Innovative Areas 2905: JP17H06358, JP17H06361 and JP17H06364, 
JSPS Core-to-Core Program A. Advanced Research Networks, 
JSPS Grant-in-Aid for Scientific Research (S) 17H06133, 
the joint research program of the Institute for Cosmic Ray Research, University of Tokyo
in Japan,
National Research Foundation (NRF) and Computing Infrastructure Project of KISTI-GSDC in Korea, 
Academia Sinica (AS), AS Grid Center (ASGC) and the Ministry of Science and Technology (MoST) in Taiwan under grants including AS-CDA-105-M06, the LIGO project, and the Virgo project.
This paper carries JGW Document Number JGW-P1910218.\\
　Jun'ya Kume is supported by a research program of the Leading Graduate Course for Frontiers of Mathematical Sciences and Physics (FMSP). This work was partially supported by JSPS KAKENHI, Grant-in-Aid for Scientific Research No.\ 15H02082 (Jun'ichi Yokoyama, Yosuke Itoh, Toyokazu Sekiguchi).\\

\end{document}